\documentclass[runningheads]{llncs}
\usepackage{times}
\usepackage{soul}
\usepackage[utf8]{inputenc}
\usepackage{graphicx}
\graphicspath{{figure/}}
\usepackage{latexsym}
\usepackage{subfigure}
\usepackage{booktabs} 
\usepackage{multirow}
\usepackage{bm}
\usepackage{mathrsfs}

\usepackage{url}
\usepackage{amsmath}
\usepackage{amssymb}
\usepackage{amsfonts}
\usepackage[misc]{ifsym}
\usepackage[linesnumbered,ruled,lined,commentsnumbered]{algorithm2e}

\begin{document}
\title{DyHGCN: A Dynamic Heterogeneous Graph Convolutional Network to Learn Users' Dynamic Preferences for Information Diffusion Prediction}
%
%
\author{
	Chunyuan Yuan\inst{1,2} \and
	Jiacheng Li\inst{1,2} \and 
	Wei Zhou\inst{1,*} \and 
	Yijun Lu\inst{3} \and
	Xiaodan Zhang\inst{1} \and 
	Songlin Hu\inst{1,2}
}
%
%
\institute{
	Institute of Information Engineering, Chinese Academy of Sciences \and
	School of Cyber Security, University of Chinese Academy of Sciences 
	\email{\{yuanchunyuan,lijiacheng,zhouwei,zhangxiaodan,husonglin\}@iie.ac.cn}  \and
	Alibaba Cloud Computing Co. Ltd.\\
	\email{\{yijun.lyj\}@alibaba-inc.com}
}

\maketitle              
%


\begin{abstract} 
	Information\let\thefootnote\relax\footnotetext{* Corresponding author.} diffusion prediction is a fundamental task for understanding the information propagation process. It has wide applications in such as misinformation spreading prediction and malicious account detection. Previous works either concentrate on utilizing the context of a single diffusion sequence or using the social network among users for information diffusion prediction. However, the diffusion paths of different messages naturally constitute a dynamic diffusion graph. For one thing, previous works cannot jointly utilize both the social network and diffusion graph for prediction, which is insufficient to model the complexity of the diffusion process and results in unsatisfactory prediction performance. For another, they cannot learn users' dynamic preferences. Intuitively, users' preferences are changing as time goes on and users' personal preference determines whether the user will repost the information. Thus, it is beneficial to consider users' dynamic preferences in information diffusion prediction. 
	
	In this paper, we propose a novel dynamic heterogeneous graph convolutional network (DyHGCN) to jointly learn the structural characteristics of the social graph and dynamic diffusion graph. Then, we encode the temporal information into the heterogeneous graph to learn the users' dynamic preferences. Finally, we apply multi-head attention to capture the context-dependency of the current diffusion path to facilitate the information diffusion prediction task. Experimental results show that DyHGCN significantly outperforms the state-of-the-art models on three public datasets, which shows the effectiveness of the proposed model. 
	
\end{abstract}


\section{Introduction}
Online social media has become an indispensable part of our daily life, on which people can deliver or repost interesting news easily. The information diffusion prediction task aims at studying how information spread among users and predicting the future infected user. The modeling and prediction of the information diffusion process play an important role in many real-world applications, such as predicting social influence~\cite{Qiu2018deepinf}, analyzing how misinformation spreads~\cite{tambuscio2015fact,rumor_yuan_2019} and detecting malicious accounts~\cite{liu2018heterogeneous,yuan2019learning}.

%
%

Previous studies either concentrate on the utilization of the diffusion sequence~\cite{yang2010modeling,du2016recurrent,wang2017topological,wang2017cascade,yang2018neural} or using the social network among users for diffusion prediction~\cite{guille2012predictive,Wang2018SNIDSA,yang2019multi}. Some studies~\cite{wang2017topological,wang2017cascade,islam2018deepdiffuse,yang2018neural} proposed the diffusion path based models to learn user representation from the past diffusion records. For example, TopoLSTM~\cite{wang2017topological} extended the standards LSTM model to learn the chain structure of the information diffusion sequence. CYAN-RNN~\cite{wang2017cascade} modeled the diffusion path as a tree structure and attention-based RNN to capture the cross-dependence based on the observed sequence. The diffusion path can reflect the information trends, so these models can achieve success in formulating the observed sequence.


Apart from utilizing the diffusion path, some studies apply the social network among users to facilitate information diffusion predication. An intuition behind it is that people have some common interests with their friends~\cite{yang2015rain}. If their friends repost the information, they have a higher probability to repost it. Based on this assumption, many recent studies~\cite{yang2015rain,bourigault2016representation,wang2018attention,yang2019multi} exploit the structure of the social network to learn the social influence among users for improving prediction performance.


However, existing methods including state-of-the-art models~\cite{Wang2018SNIDSA,yang2019multi} do not consider two important aspects: For one thing, they cannot jointly utilize both the social network and diffusion graph for prediction, which is insufficient to model the complexity of the diffusion process and results in unsatisfactory prediction performance. For another, they cannot learn users' dynamic preferences. Intuitively, users' preference is changing as time goes on and users' personal preference influences information diffusion. Thus, it is beneficial to consider the user's dynamic preference, which can be reflected by the dynamic diffusion structure at different points of diffusion time.  


To take advantage of these aspects, we propose a novel dynamic heterogeneous graph convolutional network (DyHGCN) to utilize both the social network and dynamic diffusion graph for prediction. Firstly, we design a heterogeneous graph algorithm to learn the representation of the social network and diffusion relations. Then, we encode the temporal information into the heterogeneous representation to learn the users' dynamic preference. Finally, we capture the context-dependency of the current diffusion path to solve the information diffusion prediction problem.

The main contributions of this paper can be summarized as follows:
\begin{itemize}
\item We design a dynamic heterogeneous graph convolutional network (DyHGCN) to jointly model users' social graph and diffusion graph for learning complex diffusion processes.

\item We encode the temporal information into the heterogeneous representation to learn the users' dynamic preferences. As far as we know, it is the first work to utilize users' dynamic preferences for information diffusion prediction.

\item Experimental results suggest that DyHGCN outperforms the state-of-the-art models on three public datasets, which shows the effectiveness and efficiency of DyHGCN. 
\end{itemize}




\section{Related Work} \label{relatedwork}
Current information diffusion prediction methods can be categorized into two categories: diffusion path based methods and social network based methods.

\subsection{Diffusion Path based Methods.}
The diffusion path based methods infer the interpersonal influence based on given observed diffusion sequences. Early work assumed that there is a prior diffusion model in the information diffusion process, such as the independent cascade model~\cite{kempe2003maximizing} or linear threshold model~\cite{granovetter1978threshold}. Although these models~\cite{kempe2003maximizing,saito2009learning} achieve success in formulating the implicit influence relations between users, the effectiveness of these methods relies on the hypothesis of prior information diffusion model, which is hard to specify or verify in practice~\cite{wang2017cascade}. 

With the development of neural network models, such as the recurrent neural network (RNN) and convolutional neural network (CNN),  some studies~\cite{gomez2012inferring,saito2011learning,yang2018neural} apply deep learning to automatically learn a representation of the underlying path from the past diffusion sequence for diffusion prediction, without requiring an explicit underlying diffusion model. For example, TopoLSTM~\cite{wang2017topological} extended the standards LSTM model to learn the information diffusion path to generate a topology-aware node embedding. DeepDiffuse~\cite{islam2018deepdiffuse} employed embedding technique and attention model to utilize the infection timestamp information. The model can predict when and who is going to be infected in a social media based on previously observed cascade sequence. NDM~\cite{yang2018neural} built a microscope cascade model based on self-attention and convolution neural networks to alleviate the long-term dependency problem. 

Most diffusion path based methods treat the problem as a sequence prediction task, which aims at predicting diffusion users sequentially and explore how historical diffusion sequences affect future diffusion trends. However, social relations among users are one of the critical channels of information diffusion, which is not applied in these methods. Thus, it is hard to accurately identify and predict the direction of information flow without considering the structure of social networks.



%







\subsection{Social Graph based Methods.}
Apart from utilizing the diffusion path, some studies leverage the structure of the social network for diffusion prediction. An intuition behind it is that people have some common interests with their friends~\cite{yang2015rain}. If their friends repost the news or microblogs, they have a higher probability to repost it too. Based on this assumption, many previous studies~\cite{bourigault2016representation,li2017modeling,zhang2018cosine,wang2018attention,yang2019multi,yang2015rain} have been exploring to improve prediction performance from the view of social relations. For example, \cite{yang2015rain} studied the interplay between users' social roles and their influence on information diffusion. They proposed a role-aware information diffusion model that integrated social role recognition and diffusion modeling into a unified framework. \cite{Wang2018SNIDSA} explored both the sequential nature of an information diffusion process and structural characteristics of the user connection graph and employed a RNN-based framework to model the historical sequential diffusion. \cite{yang2019multi} proposed a multi-scale diffusion prediction model based on reinforcement learning. The model incorporated the macroscopic diffusion size information into the RNN-based microscopic diffusion model. 


However, these graph based methods mainly focus on the current diffusion sequence and ignore the diffusion path of other messages in the meantime, which cannot capture global reposting relations. Thus, it is insufficient to model the complexity of the diffusion processes. Different from social graph based models, we jointly learn the global structure of the social graph and dynamic diffusion graph. Moreover, our model considers dynamic individual preference based on this heterogeneous graph.

\section{Problem Formulation} \label{preliminary}
Suppose that a collection of message $\mathcal{D}$ will be propagated among a set of users  $\mathcal{U}$. In this paper, we regard a piece of message as a document. An explicit way to describe an information diffusion process can be viewed as a successive activation of nodes indicating when people share or repost a document. Most often, real cascades are recorded as single-chain structure sequences. 

\begin{figure}
	\centering
	\includegraphics[width=\linewidth]{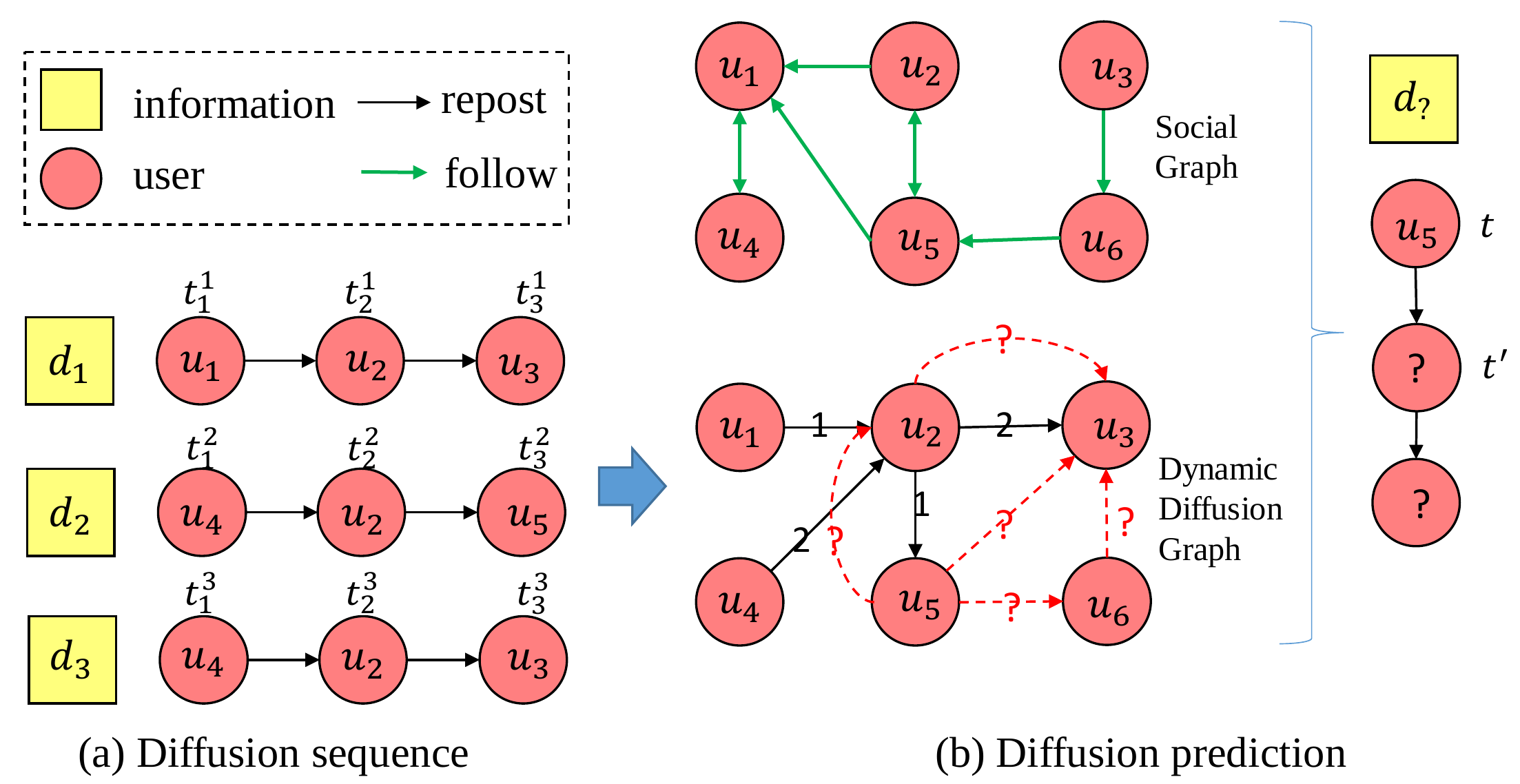}
	\caption{
		(a) An example of diffusion process of documents $d_1$, $d_2$, $d_3$ (marked as yellow squares). The edges denote that the user (marked as red circles) reposted a document at a certain timestamp. (b) Illustration of the information diffusion prediction task. The red dashed lines indicate possible reposting behavior and potential activation users.
	}
	\label{fig1}
\end{figure}

The diffusion process of document $d_m$ is recorded as a sequence of reposting behaviors $\mathcal{S}^{m}=\{s^{m}_1, s^m_2, \dots, s_{N_c}^{m} \}$, where $N_c$ is the cascade number for message ${d_m}$. $N_c$ is the maximum length of diffusion sequence. The reposting behavior $s_k^m = \{ (u_k^{m}, t_k^{m}) | u_k^m \in \mathcal{U}, t_k^{m}\in [0, +\infty) \} $ is a tuple, referring to that user $u_k^m$ reposted the message $d_m$ at a certain timestamp $t_k^{m}$. As shown in Fig.~\ref{fig1}(a), for document $d_1$ is recorded as $\{(u_1, t^1_1),(u_2, t^1_2),(u_3, t^1_3), \ldots \}$ order by timestamp.

Given the observed diffusion traces, the target of the information diffusion prediction task is to predict the diffusion behavior at a future timestamp $t'$. As shown in Fig.~\ref{fig1}(b), we have known that user $u_5$ delivers a piece of information $d_?$ and we should predict which users would be interested in it and will repost it in the future timestamp $t'$. The diffusion probability can be formulated as $P(s^m_{t'}|\mathcal{S}_{t})$, where $t' > t$. 

In this paper, we jointly encode the users' social graph and the diffusion graph to solve the diffusion prediction problem. The motivation is that both graphs provide different useful information for diffusion prediction and combining them will make predictions more accurate. As shown in Fig.~\ref{fig1}(b), both $u_2$ and $u_6$ have higher probability to repost message from $u_5$ than $u_3$, because we can observe that $u_2$ and $u_6$ follow $u_5$ from social graph, whereas $u_3$ did not directly follow $u_5$. However, the message would be likely to be propagated to $u_3$ by $u_2$ or $u_6$, because $u_3$ follows $u_6$ and $u_3$ has twice reposting records from $u_2$. From this example, we can see that it is beneficial to combine both graphs for the diffusion prediction problem.

\section{Framework}  \label{model}
\begin{figure}
	\centering
	\includegraphics[scale=0.5]{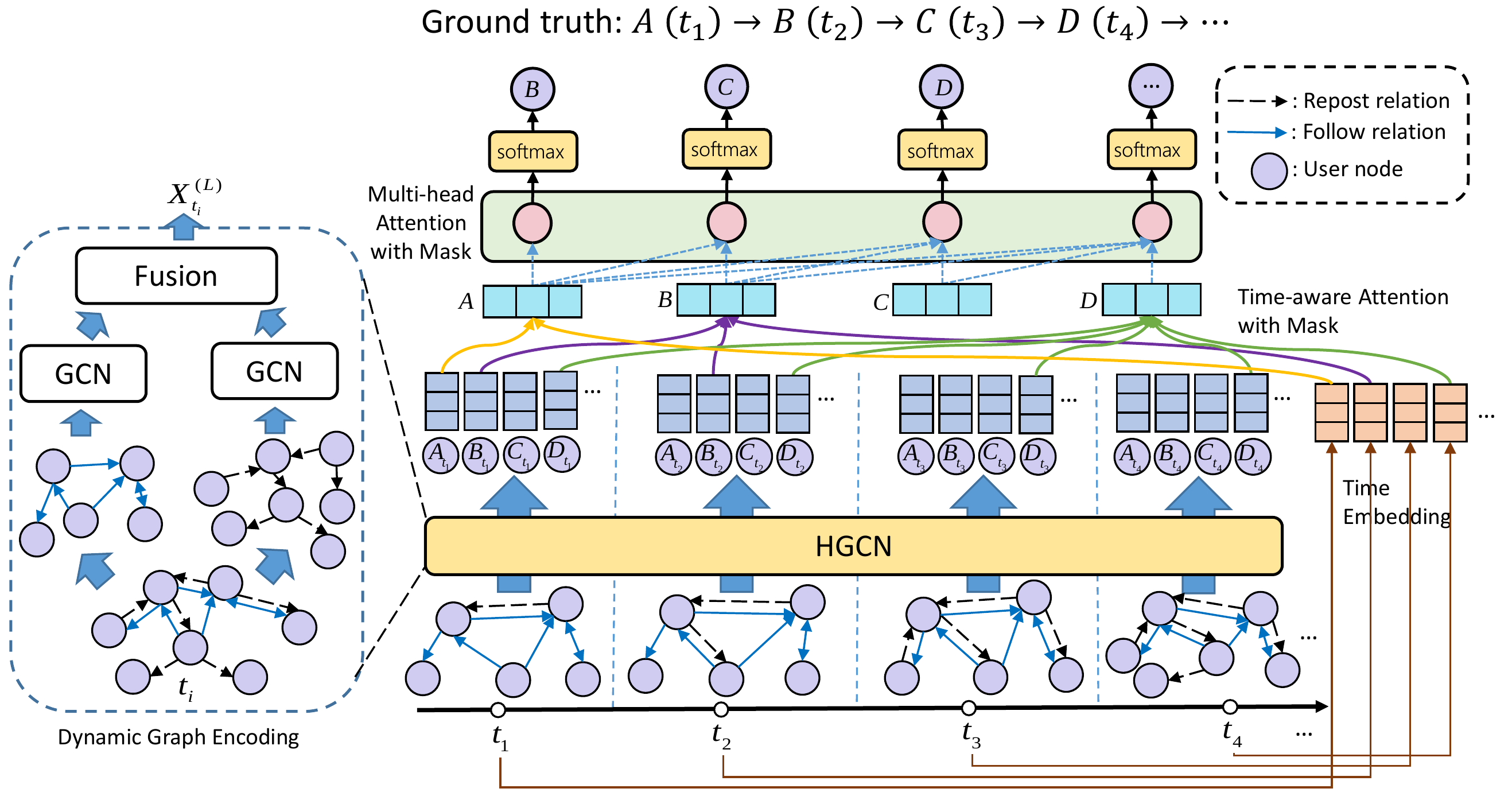}
	\caption{The architecture of dynamic heterogeneous graph convolutional network.}
	\label{fig:framework}
\end{figure}

In this section, we will introduce the DyHGCN, a deep learning-based model with three stages, to elaborately learn the individual dynamic preference and the neighbors' influence. The overview architecture of the proposed model is shown in Fig.~\ref{fig:framework}. Firstly, we construct a heterogeneous social and diffusion graph. Then, we design the dynamic heterogeneous graph to learn the node representations of the graph with users' dynamic preferences. Finally, we combine these representations to predict future infected users.

\subsection{Heterogeneous Graph Construction}  
Intuitively, people would repost the message or microblog when they are interested in it. Users usually follow someone if they like his or her microblog. Therefore, social relations in the social graph would be useful to predict whether the user will repost the message. Furthermore, we can analyze the history of reposting behaviors at different diffusion period. In this way, we can capture the dynamic changes in users' preferences. Based on these motivations, we propose to jointly model the social relations and dynamic repost relations to learn better user representations for information diffusion prediction. 

In this paper, we utilize the social relations among users to construct a user social graph $\mathcal{G}^f$, which is a directed and unweighted graph. Then, we split the historical diffusion timeline into $n$ time intervals. At each time interval $t_i$, we use the repost relations among users to construct a diffusion graph $\mathcal{G}_{t_i}^r$, which is a directed and weighted graph.

\subsection{Heterogeneous Graph Convolutional Network (HGCN)}
As shown in the left part of Fig.~\ref{fig:framework}, the heterogeneous graph has one type of node (user node) and two types of relations: follow relation and repost relation. At time interval $t_i, i \in [1, n]$, we use both relations to construct the adjacency matrix $\mathbf{A} = \{\mathbf{A}^F, \mathbf{A}_{t_i}^R\}$. $\mathbf{A}^F \in \mathbb{R}^{|U| \times |U|}$ is the adjacency matrix extracted from social relations, and $\mathbf{A}_{t_i}^R \in \mathbb{R}^{|U| \times |U|}$ is extracted from the repost relations. $|U|$ denotes the amount of users. 

For each kind of relation, we apply a multi-layer graph convolutional network (GCN)~\cite{kipf2017semi} to learn the node representation from the graph. The layer-wise propagation rule can be defined as follows: 
\begin{equation}
    \begin{split}
        & \mathbf{X}_F^{(l+1)} = \sigma(\mathbf{A}^F \mathbf{X}^{(l)} \mathbf{W}_F^{(l)} ) \,, \\
        & \mathbf{X}_R^{(l+1)} = \sigma(\mathbf{A}_{t_i}^R (\mathbf{X}^{(l)} + \mathbf{t}_i) \mathbf{W}_R^{(l)} ) \,, \\
    \end{split}
\end{equation}
where $\mathbf{X}^{(0)} \in \mathbb{R}^{|U| \times d}$ is randomly initialized user embeddings by a normal distribution~\cite{glorot2010understanding}, and $\mathbf{W}_F^{(l)}, \mathbf{W}_R^{(l)} \in \mathbb{R}^{d \times d}$ are learnable parameters. $\mathbf{t}_i \in \mathbb{R}^{d}$ is the time interval embedding initialized by a random distribution. $d$ is the dimensionality of user embeddings. $\sigma(\cdot)$ is the $\mathbf{ReLU}(x) = \max (0, x)$ activation function. $l$ denotes the layers of GCN.

We can obtain the user representations $\mathbf{X}_F^{(l+1)} \in \mathbb{R}^{|U| \times d}$ from follow relations and $\mathbf{X}_R^{(l+1)} \in \mathbb{R}^{|U| \times d}$ from repost relations. To fuse both relations for generating better user representation, we apply a heuristic strategy~\cite{mou2016natural} to make both $\mathbf{X}_F^{(l+1)}$ and $\mathbf{X}_R^{(l+1)}$ interact with each other:
\begin{equation}
    \begin{split}
        & \mathbf{X}_{FR}^{(l+1)} = [\mathbf{X}_F^{(l+1)}; \mathbf{X}_R^{(l+1)}; \mathbf{X}_F^{(l+1)} \odot \mathbf{X}_R^{(l+1)}; \mathbf{X}_F^{(l+1)}-\mathbf{X}_R^{(l+1)}] \,, \\
        & \mathbf{X}_{t_i}^{(l+1)} = \mathbf{X}^{(l+1)}_{FR} \mathbf{W}_1 \,, \\
    \end{split}
    \label{fuse_node_rep}
\end{equation}
where $\odot$ denotes the element-wise product and $\mathbf{W}_1 \in \mathbb{R}^{4d \times d}$ is a learnable parameter. $\mathbf{X}_{t_i}^{(l+1)}$ is the learnt user representation at time $t_i$.

\subsection{Dynamic Graph Encoding}
As discussed above, users' dynamic preference is important for diffusion prediction. In this section, we will describe how to learn user representations from the dynamic graphs at different time intervals. 

\begin{algorithm}[!htb] 
	\caption{The dynamic graph encoding algorithm.} 
	\label{dynamic_graph_encoding} 
	\footnotesize 
	\KwIn{
		\indent The heterogeneous graph $\mathcal{G(V, E)}$; \\
		\indent Time intervals $T = \{t_1, t_2, \ldots, t_n \}$; \\
		\indent Adjacency matrix $\mathbf{A}^F, and [\mathbf{A}_{t_1}^R, \mathbf{A}_{t_2}^R, \ldots, \mathbf{A}_{t_n}^R]$.
	}
    
	\KwOut{A list of user representations at all time intervals.}
	
	\For{each $t_i \in T$} 
	{
         Construct adjacency matrix $A^R_{t_i}$ from diffusing sub-graph. \\
         Construct adjacency matrix $A^F$ from social sub-graph. \\
         \For{$l = 1, \ldots, L$} {
            $\mathbf{X}_F^{(l+1)} = \sigma(\mathbf{A}^F X^{(l)} W_F^{(l)} )$ ;  \\
            $\mathbf{X}_R^{(l+1)} = \sigma(\mathbf{A}_{t_i}^R (X^{(l)} + \mathbf{t}_i) W_R^{(l)})$ ; \\
            Fuse $\mathbf{X}_F^{(l+1)}$ and $\mathbf{X}_R^{(l+1)}$ to form $\mathbf{X}_{t_i}^{(l+1)}$ by Equation~(\ref{fuse_node_rep}).
         }
         Collect user representation $\mathbf{X}_{t_i}^{(L)}$ from HGCN.
	}
	\Return $\mathbf{X}_{t_1}^{(L)}, \mathbf{X}_{t_2}^{(L)}, \ldots, \mathbf{X}_{t_n}^{(L)}$
\end{algorithm}

The overall process of the dynamic graph encoding algorithm is shown in Algorithm~\ref{dynamic_graph_encoding}. Firstly, we split the historical diffusion timeline as $n$ intervals to construct $n$ dynamic heterogeneous graphs. Then, we apply the heterogeneous graph convolutional network to learn the user embeddings at each time interval. Finally, we collect all user representations and send them to the next stage. 




\subsection{Time-aware Attention}
After the above procedures, we obtain user representations from different heterogeneous graph snapshots of different time intervals. Then, we can generate the final user representation by fusing these user representations of different time intervals. In this subsection, we design two kinds of strategies to produce final user representations.

\subsubsection{Hard Selection Strategy} 

\

\noindent For every user in the diffusion trace, they all have a timestamp during repost. We can directly determine to which time interval a given timestamp belongs. Then we use the user embedding of that time interval as final user embedding. For example, given user id $u$, we look up user representation from all user representations $[\mathbf{X}_{t_1}^{(L)}, \mathbf{X}_{t_2}^{(L)}, \ldots, \mathbf{X}_{t_n}^{(L)}]$. We will obtain $n$ user representations $[\mathbf{u}_{t_1}, \mathbf{u}_{t_2}, \ldots, \mathbf{u}_{t_n}]$. Supposing user $u$ reposts the information at timestamp $t'$ and $t' \in [t_3, t_4)$, then we use $\mathbf{u}_{t_3}$ as final representations of user $u$.

\subsubsection{Soft Selection Strategy}

\

\noindent The hard selection strategy only uses the user representation belonging to the repost time interval, which can not fully utilize the user representations produced by historical information. Thus, we design a time-aware attention module to fuse the historical user representations as final user representation. 

Specifically, given user id $u$, we look up user representation from all user representations $[\mathbf{X}_{t_1}^{(L)}, \mathbf{X}_{t_2}^{(L)}, \ldots, \mathbf{X}_{t_n}^{(L)}]$, and obtain user representations $\mathbf{U}_t = [\mathbf{u}_{t_1}, \mathbf{u}_{t_2}, \ldots, \mathbf{u}_{t_n}] \in \mathbb{R}^{n \times d}$. Supposing user $u$ reposts the information at timestamp $t'$ and $t' \in [t_3, t_4)$, then we define the time-aware attention as follows:
\begin{equation}
    \begin{split}
        & \mathbf{t}' = \mathbf{Lookup}(t_3)  \,,  \\
        & \mathbf{\alpha} = \mathbf{softmax}(\frac{\mathbf{U}_{t}^T \mathbf{t}'}{\sqrt{d}} + \mathbf{m})   \,,  \\
        & \widetilde{\mathbf{u}} = \sum_{i=1}^{n}{\mathbf{\alpha}_i \mathbf{U}_{t_i} }
    \end{split}
\end{equation}
where $\mathbf{m}_j = \begin{cases}
0& \text{$t' \geq t_j$,}\\
-\infty& \text{otherwise.}
\end{cases} $ is a mask matrix and $\mathbf{m} \in \mathbb{R}^{n}$. When $\mathbf{m}_j=-\infty$, the softmax function results in a zero attention weight, which can switch off the attention when $t' < t_j$ to avoid leaking labels of future timestamp. $\mathbf{Lookup}(\cdot)$ function is applied to transform the time interval id into time embedding. The time embeddings are initialized by a normal distribution~\cite{glorot2010understanding}. $\widetilde{\mathbf{u}}$ is the final representation of user $u$.

\subsection{Information Diffusion Prediction}
To capture the context-dependency information, we can apply the learned user representations to construct the current diffusion sequence $\widetilde{\mathbf{U}} = [\widetilde{\mathbf{u}}_A, \widetilde{\mathbf{u}}_B,  \widetilde{\mathbf{u}}_C, \ldots]$ for future diffusion prediction. Instead of using a recurrent neural network (RNN) to model the current diffusion sequence, we apply masked multi-head self-attention module~\cite{vaswani2017attention} to parallelly attend to each other for context encoding. Compared with RNN, a multi-head attention module is much faster and easier to learn the context information. It is worth noting that we also apply mask matrix as before to mask the future information to avoid leaking labels. The process can be formulated as follows:
\begin{equation}
    \begin{split}
        & \mathbf{Attention}(\mathbf{Q}, \mathbf{K}, \mathbf{V}) = \mathbf{softmax}\left( \frac{\mathbf{Q} \mathbf{K}^T}{\sqrt{d}_k} + \mathbf{M} \right) \mathbf{V} \,, \\
        & \mathbf{h}_i = \mathbf{Attention}\left(\widetilde{\mathbf{U}} \mathbf{W}_{i}^{Q}, \widetilde{\mathbf{U}} \mathbf{W}_{i}^{K}, \widetilde{\mathbf{U}} \mathbf{W}_{i}^{V}\right)  \,,  \\
        & \mathbf{Z} = [\mathbf{h}_{1}; \mathbf{h}_{2}; \ldots; \mathbf{h}_{H}] \mathbf{W}^{O}
    \end{split}
\end{equation}
where $W_{i}^{Q}, W_{i}^{K}, W_{i}^{V} \in \mathbb{R}^{d \times d_{k} }$ and $W^{O}\in \mathbb{R}^{H d_{k}\times d_{Q}}$; $d_k=d / H$; $H$ is the number of heads of attention module. The mask matrix $\mathbf{M}$, which is defined as:
\begin{equation}
    \begin{split}
        & \mathbf{M}_{ij} = \begin{cases}
                                0& \text{$i \le j$,}\\
                                -\infty& \text{otherwise.}
                            \end{cases} 
    \end{split}
\end{equation} is used to switch off attention weights of the future time step.

We obtain user representations $\mathbf{Z} \in \mathbb{R}^{L \times d}$ on $L$ diffusion time steps. Then, we use two layers fully-connected neural network to compute the diffusion probability:
\begin{equation}
 \hat{\mathbf{y}} = \mathbf{W}_3 \mathbf{ReLU}(\mathbf{W}_2 \mathbf{Z}^T + \mathbf{b}_1) + \mathbf{b}_2  \,,
\end{equation}
where $\hat{\mathbf{y}} \in \mathbb{R}^{L \times |U|}$, and $\mathbf{W}_2 \in \mathbb{R}^{d \times d}, \mathbf{W}_3 \in \mathbb{R}^{|U| \times d}, \mathbf{b}_1, \mathbf{b}_2$ are the learnable parameters.

Finally, we apply the cross entropy loss as the objective function, which is formulated as: 
\begin{equation}
  \mathcal{J}(\mathbf{\theta}) = -\sum_{i=2}^{L} \sum_{j=1}^{|U|} \mathbf{y}_{ij} \log(\hat{\mathbf{y}}_{ij})
\end{equation} 
where $\mathbf{y}_{ij} = 1$ denotes that the diffusion behavior happened, otherwise $\mathbf{y}_{ij} = 0$. $\theta$ denotes all parameters needed to be learned in the model. The parameters are updated by Adam optimizer with mini-batch.






\section{Experiments} \label{experiment}

\subsection{Datasets}
Following the previous studies~\cite{Wang2018SNIDSA,yang2019multi}, we conduct experiments on three public datasets to quantitatively evaluate the proposed model. The detailed statistics are presented in Table~\ref{tab:datastatistics}. \textit{\#Links} denotes the amount of follow relations of users in the social network. \textit{\#Cascades} denotes the amount of diffusion sequence in the dataset. \textit{Avg. Length} indicates the average length of the information diffusion sequence.

\begin{table}[!htbp]
    \centering
	\caption{
	    Statistics of the Twitter, Douban, and Memetracker datasets. 
	}
	\label{tab:datastatistics}
	\setlength{\tabcolsep}{3mm}{
		\begin{tabular}{p{2cm}|ccc}
    		 \toprule[1.5pt]
    		 Datasets       & Twitter & Douban & Memetracker \\
    		 \midrule
    		 \# Users       & 12,627     &  23,123   & 4,709 \\
    		 \# Links       & 309,631    &  348,280  & - \\
    		 \# Cascades    & 3,442      &  10,602   & 12,661 \\
    		 Avg. Length    & 32.60      &  27.14    & 16.24 \\
    		 \bottomrule[1.5pt]
		\end{tabular}
	}
\end{table}

\textbf{Twitter}\footnote{http://www.twitter.com} dataset~\cite{hodas2014simple} records the tweets containing URLs during October 2010. Each URL is interpreted as an information item spreading among users. The social relation of users is the follow relation on Twitter.

\textbf{Douban}\footnote{http://www.douban.com} dataset~\cite{zhong2012comsoc} is collected from a social website where users can update their book reading statuses and follow the statuses of other users. Each book or movie is considered an information item and a user is infected if she reads or watches it. The social relation of users is the co-occurrence relation. If two users take part in the same discussion more than 20 times, they are considered a friend.

\textbf{Memetracker} dataset~\cite{leskovec2009meme} collects millions of news stories and blog posts from online websites and tracks the most frequent quotes and phrases, i.e. memes, to analyze the migration of memes among people. Each meme is regarded as an information item and each URL of websites is treated as a user. Note that this dataset has no underlying social graph. 

As in previous studies~\cite{Wang2018SNIDSA,yang2019multi}, we randomly sample 80\% of cascades for training, 10\% for validation and the rest 10\% for test. The statistics of datasets are listed in Table~\ref{tab:datastatistics}.

\subsection{Comparison Methods}
The comparison baselines can be divided into diffusion path based models and social network based models. To evaluate the effectiveness of DyHGCN, we use five very recent models as baselines for a thorough comparison. The models are shown as follows:

\subsubsection{Based on diffusion path}

\begin{itemize}
\item \textbf{TopoLSTM}~\cite{wang2017topological}: models the information diffusion path as a dynamic directed acyclic graph and extends the standard LSTM model to learn a topology-aware user embedding for diffusion prediction. 

\item \textbf{DeepDiffuse}~\cite{islam2018deepdiffuse}: employs the embedding technique and attention model to utilize the infection timestamp information. The model can predict when and who is going to be infected in a social network based on previously observed cascade sequence.


\item \textbf{NDM}~\cite{yang2018neural}: builds a microscopic cascade model based on the self-attention mechanism and convolution neural networks to alleviate the long-term dependency problem.
\end{itemize}

\subsubsection{Based on social network}

\begin{itemize}
\item \textbf{SNIDSA}~\cite{Wang2018SNIDSA}: is a sequential neural network with structure attention to model information diffusion. The recurrent neural network framework is employed to model the sequential information. The attention mechanism is incorporated to capture the structural dependency among users. A gating mechanism is developed to integrate sequential and structural information. 

\item \textbf{FOREST}~\cite{yang2019multi}: is a multi-scale diffusion prediction model based on reinforcement learning. The model incorporates the macroscopic diffusion size information into the RNN-based microscopic diffusion model. It is the latest sequential model and achieves state-of-the-art performance.
\end{itemize}






\noindent\textbf{Our methods (DyHGCN-H, DyHGCN-S):} DyHGCN-H is the model with hard selection strategy, and DyHGCN-S is the model with soft selection strategy (time-aware attention). 


\subsection{Evaluation Metrics and Parameter Settings.}
Following the settings of previous studies~\cite{wang2017cascade,Wang2018SNIDSA,yang2019multi}, we consider the next infected user prediction as a retrieval task by ranking the uninfected users by their infection probabilities. We evaluate the performance of DyHGCN with state-of-the-art baselines in terms of Mean Average Precision (MAP) on top k (Map@k) and HITS scores on top k (Hits@k).

Our model is implemented by PyTorch~\cite{paszke2017automatic}. The parameters are updated by Adam algorithm~\cite{kingma2014adam} and the parameters of Adam, $\beta_1$ and $\beta_2$ are 0.9 and 0.999 respectively. The learning rate is initialized as 1e-3. The batch size of the training set is set to 16. The dimensionality of user embedding and temporal interval embedding are set to $d=64$. We use two layer of GCN tp learn the graph structure. The kernel size is set to 128. The number of heads in multi-head attention $H$ is chosen from $\{2, 4, 6, 8, 10, 12, 14, 16, 18, 20\}$ and finally set to 14. We split the dynamic diffusion graph into $n$ time intervals, where $n = \{1, 2, 4, 6, 8, 10, 12, 14, 16, 18, 20 \}$. Finally, we use $n=8$ in the experiment. We select the best parameter configuration based on performance on the validation set and evaluate the configuration on the test set.


\subsection{Experimental Results}
We evaluate the effectiveness of DyHGCN on three public datasets for information diffusion prediction task. Table~\ref{exp_results_on_twitter}, \ref{exp_results_on_douban} and \ref{exp_results_on_memetracker} show the performance of all methods.

\begin{table}
	\centering
	\caption{
		Experimental results on Twitter dataset (\%). All experimental results of baselines are cited from paper~\cite{yang2019multi}. FOREST~\cite{yang2019multi} is the state-of-the-art model until this submission. Improvements of DyHGCN are statistically significant with $p < 0.01$ on paired t-test.
	}
	\setlength{\tabcolsep}{2mm}{
		\begin{tabular}{p{2cm}|ccc|ccc}
			\toprule[1.2pt]
			\multirow{2}[2]{*}{Models} & \multicolumn{6}{c}{Twitter} \\
			& $hits@10$ & $hits@50$ & $hits@100$ & $map@10$ & $map@50$ & $map@100$  \\
			\midrule
			
			DeepDiffuse &  4.57&   8.80&  13.39&  3.62&  3.79& 3.85  \\
			
			TopoLSTM    &  6.51&  15.48&  23.68&  4.31&  4.67&  4.79  \\
			
			NDM         & 21.52&  32.23&  38.31&  14.30&  14.80&  14.89   \\
			
			SNIDSA & 23.37&  35.46&  43.39&  14.84&  15.40&  15.51  \\
			
			FOREST & 26.18&  40.95&  50.39&  17.21&  17.88&  18.02  \\
			
			\midrule
			DyHGCN-H & 28.48 & 47.18 & 58.48 & 16.78 & 17.63 & 17.79  \\
			DyHGCN-S & \textbf{28.98} & \textbf{47.89} & \textbf{58.85} & \textbf{17.46} & \textbf{18.30} & \textbf{18.45}    \\
			\bottomrule[1.2pt]
		\end{tabular}
		\label{exp_results_on_twitter}
	}
\end{table}

\begin{table}
	\centering
	\caption{
		Experimental results on Douban dataset (\%). All experimental results of baselines are cited from paper~\cite{yang2019multi}. FOREST~\cite{yang2019multi} is the state-of-the-art model until this submission. Improvements of DyHGCN are statistically significant with $p < 0.01$ on paired t-test.
	}
	\setlength{\tabcolsep}{2mm}{
		\begin{tabular}{p{2cm}|ccc|ccc}
			\toprule[1.2pt]
			\multirow{2}[2]{*}{Models} &  \multicolumn{6}{c}{Douban}\\
			            & $hits@10$ & $hits@50$ & $hits@100$ & $map@10$ & $map@50$ & $map@100$  \\
			\midrule
			
			DeepDiffuse &   9.02&  14.93&  19.13&  4.80&  5.07&  5.13 \\
			
			TopoLSTM    &  9.16&  14.94&  18.93&  5.00&  5.26&  5.32  \\
			
			NDM         &  10.31&   18.87&  24.02&  5.54&  5.93&  6.00   \\
			
			SNIDSA      &  11.81&   21.91&  28.37&   6.36&   6.81&  6.91  \\
			
			FOREST      &  14.16&   24.79&  31.25&   7.89&   8.38&  8.47  \\
			
			\midrule
			DyHGCN-H    & 15.69&  \textbf{28.95}&  \textbf{36.45}&  8.42&  9.03& 9.13  \\
			DyHGCN-S    & \textbf{16.34} & 28.91 & 36.13 & \textbf{9.10} & \textbf{9.67} & \textbf{9.78}  \\
			\bottomrule[1.2pt]
		\end{tabular}
		\label{exp_results_on_douban}
	}
\end{table}

\begin{table}[!h]
	\centering
	\setlength{\tabcolsep}{2mm}{
		\caption{
			Experimental results on Memetracker dataset (\%). We exclude TopoLSTM and SNIDSA for Memetracker because of the absence of underlying social graph. Improvements of DyHGCN are statistically significant with $p < 0.01$ on paired t-test.
		}
		\begin{tabular}{p{2cm}|ccc|ccc}
			\toprule[1.2pt]
			\multirow{2}[2]{*}{Models} & \multicolumn{6}{c}{Memetracker} \\
			& $hits@10$ & $hits@50$ & $hits@100$ & $map@10$ & $map@50$ & $map@100$  \\
			\midrule
			
			DeepDiffuse & 13.93 & 26.50 & 34.77 & 8.14 & 8.69 & 8.80    \\
			
			NDM & 25.44 & 42.19 & 51.14 & 13.57 &  14.33 & 14.46    \\
			
			
			FOREST & 29.43 & 47.41 & 56.77 & 16.37 & 17.21 & 17.34  \\
			
			\midrule
			DyHGCN-H & 29.63 &  \textbf{48.78} &  \textbf{58.78} & 16.33 & 17.21 & 17.36  \\
			DyHGCN-S & \textbf{29.90} & 48.30 & 58.43 & \textbf{17.64} & \textbf{18.48} & \textbf{18.63}   \\
			
			\bottomrule[1.2pt]
		\end{tabular}
		\label{exp_results_on_memetracker}
	}
\end{table}

From the table, we can see that DyHGCN (DyHGCN-H and DyHGCN-S) consistently outperforms the state-of-the-art methods by an absolute improvement of more than 5\% in terms of hits@100 and map@100 scores. Specifically, we have the following observations: 

(1) Compared with TopoLSTM, DeepDiffuse, and NDM, DyHGCN-S achieves about 5\% absolute improvement on hits@10, and over 10\% improvement on hits@100. Moreover, the prediction precision also achieves about 2\% absolute improvement. These baseline models mainly model the diffusion path as a sequence or graph structure, which ignores the social network information. However, the social network can reflect user preference. The experimental results show that it is important to consider the user social network for information diffusion prediction. 

(2) Compared with SNIDSA and FOREST, DyHGCN-S achieves over 2\% absolute improvement on hits@10, and over 5\% improvement on hits@100 on Twitter and Douban datasets. Both SNIDSA and FOREST exploit user social relations to facilitate diffusion prediction. However, when predicting the diffusion path, they only model the history diffusion path as a sequential pattern, which is insufficient to model the complex diffusion behavior and users' dynamic preference. The improvement of DyHGCN shows that it is necessary to model the diffusion path as a graph rather than a sequence or tree structure.

(3) Compared with DyHGCN-H, DyHGCN-S also shows better performance on three datasets. DyHGCN-H only uses the current state of diffusion graph to learn user embedding, which cannot capture the user's dynamic preference well. DyHGCN-S utilizes the time-aware attention module to fuse the history and current diffusion graph for producing better user representations for diffusion prediction.

\section{Further Study}  \label{furtheranalysis}

\subsection{Ablation Study}
To figure out the relative importance of every module in DyHGCN, we perform a series of ablation studies over the different parts of the model. The experimental results are presented in Table~\ref{tab:variants}. The ablation studies are conducted as following orders: 
 \begin{itemize}
 	\item \textbf{w/o time-aware attention}: Replace the time-aware attention module with a hard selection strategy. 
 	\item \textbf{w/o social graph}:  Removing the social graph convolutional network. 
 	\item \textbf{w/o diffusion graph}: Removing the diffusion graph convolutional network.
 	\item \textbf{w/o heterogeneous graph}: Removing heterogeneous graph encoding modules and randomly initializing the user representations.
 \end{itemize}

\begin{table}[!htbp]
	\centering
	\caption{Ablation study on Twitter and Douban datasets (\%)}
	\label{tab:variants}
	\setlength{\tabcolsep}{1.5mm}{
		\begin{tabular}{p{3.2cm}|ccc|ccc}
		 \toprule[1.2pt]
		 \textbf{Models} & \textbf{hits@10} & \textbf{hits@50} & \textbf{hits@100}  & \textbf{hits@10} & \textbf{hits@50} & \textbf{hits@100} \\
		 \hline
			\textbf{DyHGCN-S}          & 28.98 & 47.89 & 58.85 & 16.34 & 28.91 & 36.13 \\ 
			\hline
		    w/o time-aware attention   & 28.48 & 47.18 & 58.48  & 15.69&  28.95 & 36.45   \\
		    w/o social graph           & 27.76 & 45.31 & 56.60  & 14.28&  25.72 & 33.63  \\
		    w/o diffusion graph        & 28.27 & 46.53 & 57.22  & 14.62&  26.14 & 34.28  \\
			w/o heterogeneous graph    & 27.63 & 42.31 & 51.13  & 13.58&  23.16 & 31.15 \\
		 \bottomrule[1.2pt]
		\end{tabular}
	}
\end{table}

Table~\ref{tab:variants} shows the overall performance on several variant methods of DyHGCN. Referring to the experimental results in the table, we can observe that:

(1) When replacing the time-aware attention with a hard selection strategy, the performance drops a little compared with DyHGCN-S. The experimental results show that time-aware attention can effectively fuse the user representations produced by historical information to generate better user representation.

(2) When we remove the social graph encoding modules, the performance degrades a lot compared with DyHGCN. A similar phenomenon can be seen when removing the diffusion graph. The results indicate that both the social and repost relations encoding modules in DyHGCN are essential for information diffusion prediction. 

(3) When removing the heterogeneous graph, the performance further decays a lot compared with removing the social graph or diffusion graph. The phenomenon shows that both relations contain complementary information and combining them does help to improve the performance.

\subsection{Parameter Analysis}
In this section, we conduct some sensitivity analysis experiments of hyper-parameters on the Twitter dataset. We analyze how different choices of the hyper-parameter may affect performance. 

\begin{figure}
	\centering 
	\subfigure[Number of time intervals (\%)]{
		\label{fig:subfig:a1} 
		\includegraphics[scale=0.64]{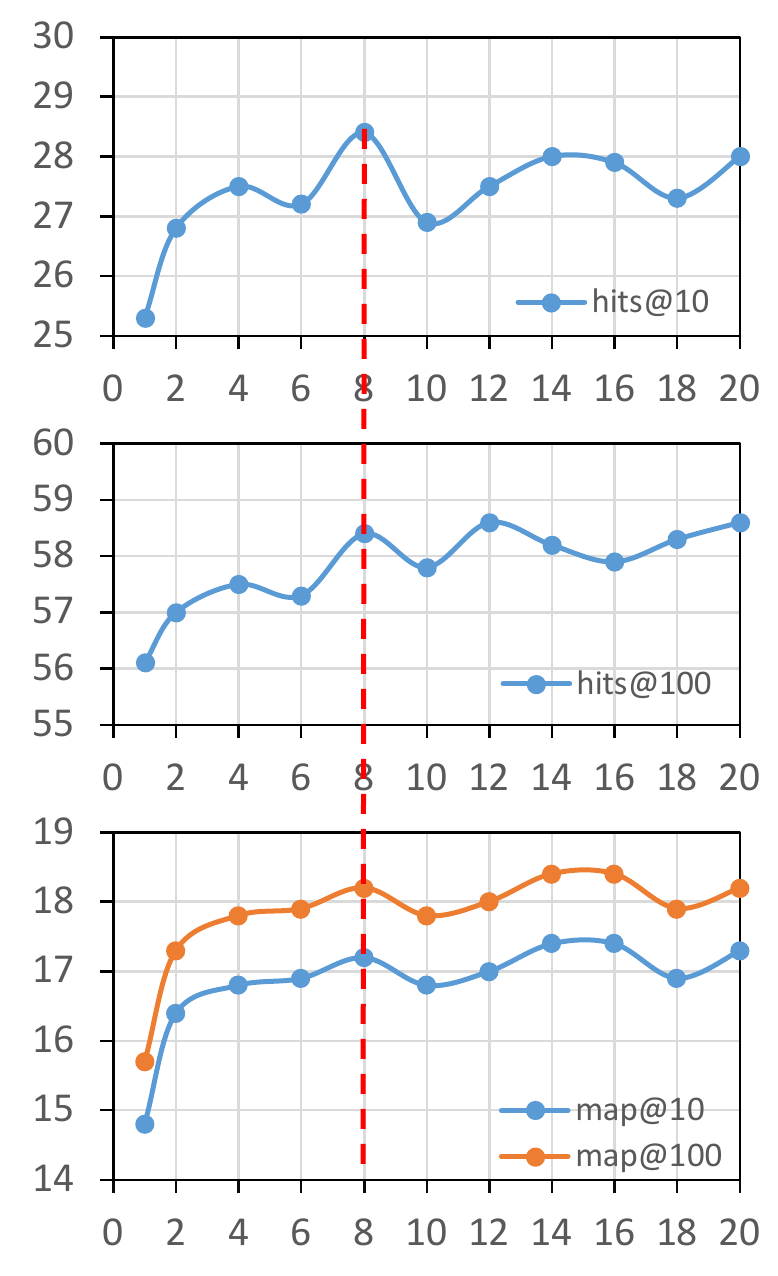}
	}
	\subfigure[Number of heads (\%)]{  
		\label{fig:subfig:b1} 
		\includegraphics[scale=0.64]{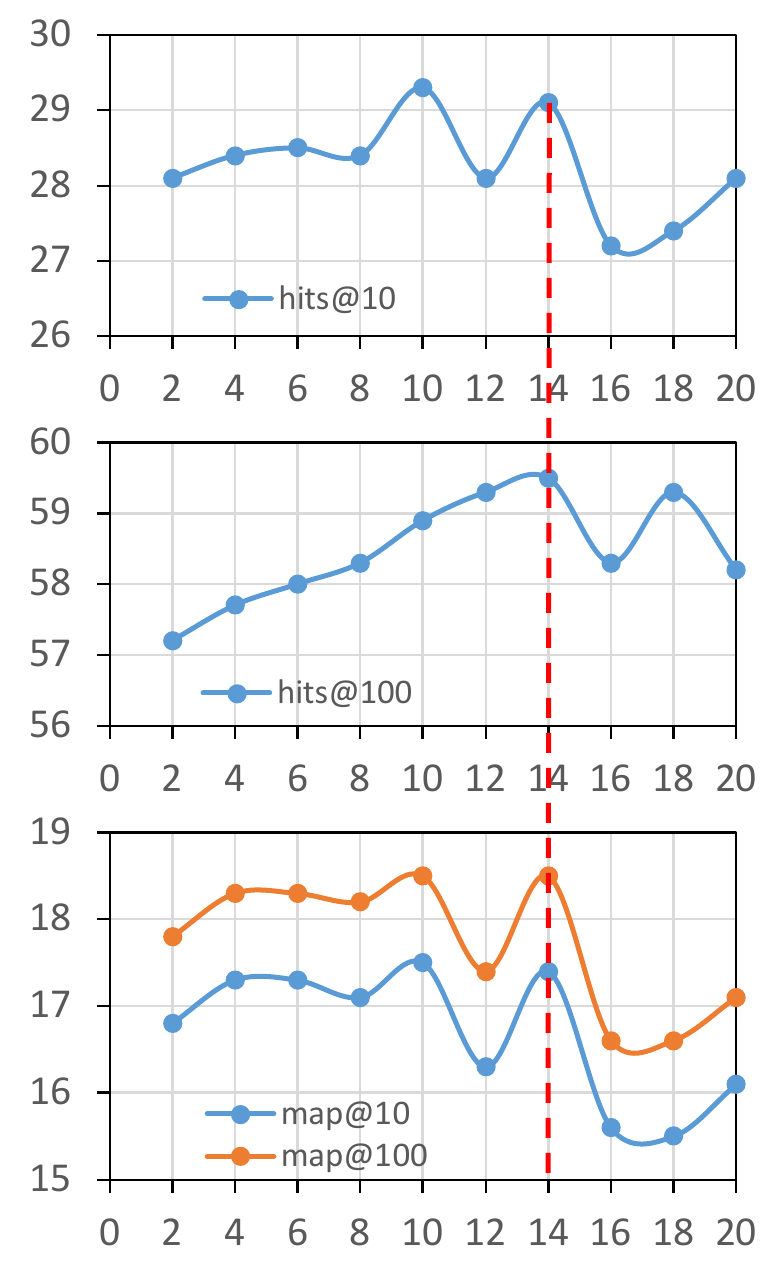}
	}
	\caption{Parameter analysis on the Twitter dataset.}
	\label{parameter_sensitivity_analysis}
\end{figure}

\textbf{Number of time intervals $n$.} In section 4.2, we split the diffusion graph into $n$ snapshots according to the diffusion timeline. The size of $n$ may affect the performance. When $n$ is larger, the diffusion graph is split many pieces and the model can learn more fine-grained changes in the dynamic graph. Referring to Fig.~\ref{fig:subfig:a1}, we can observe that : (1) Learning the dynamic characteristics of diffusion graph is helpful for information diffusion prediction, because performance is increasing when $n$ increases before $n=8$. (2) When $n$ is too large, the further improvement of performance is very limited. 

\textbf{Number of heads $H$.} From the Fig.~\ref{fig:subfig:b1}, we can see that the performance is improved a little as the increasing number of heads of multi-head attention. The model can capture more abundant information with an increase of the number of heads. However, when using too many heads, the performance drops significantly due to overfitting. We can observe that $H=14$ is the most suitable number of attention heads.

\section{Conclusion and Future Work}  \label{conclusion}
In this paper, we study the information diffusion prediction problem. To learn the dynamic preference of users for prediction, we propose a novel dynamic heterogeneous graph network to encode both the social and dynamic diffusion graph structure. We conduct experiments on three real-world datasets. The experimental results show that our model achieves significant improvements over state-of-the-art models, which shows the effectiveness and feasibility of the model for real-world applications.

For future work, we will study the text content of the diffused information, which is not applied in this work. If the users show preference about some particular topic or content, it is likely for users to repost them. So whether a user will repost the message is also determined by the content. Thus, it is worth further studying to help improve diffusion prediction performance.

%
%


\bibliographystyle{splncs04}
\bibliography{main}

\begin{thebibliography}{10}
\providecommand{\url}[1]{\texttt{#1}}
\providecommand{\urlprefix}{URL }
\providecommand{\doi}[1]{https://doi.org/#1}

\bibitem{bourigault2016representation}
Bourigault, S., Lamprier, S., Gallinari, P.: Representation learning for
  information diffusion through social networks: an embedded cascade model. In:
  Proceedings of the Ninth ACM international conference on Web Search and Data
  Mining. pp. 573--582 (2016)

\bibitem{du2016recurrent}
Du, N., Dai, H., Trivedi, R., Upadhyay, U., Gomez-Rodriguez, M., Song, L.:
  Recurrent marked temporal point processes: Embedding event history to vector.
  In: Proceedings of the 22nd ACM SIGKDD International Conference on Knowledge
  Discovery and Data Mining. pp. 1555--1564 (2016)

\bibitem{glorot2010understanding}
Glorot, X., Bengio, Y.: Understanding the difficulty of training deep
  feedforward neural networks. In: Proceedings of the thirteenth international
  conference on artificial intelligence and statistics. pp. 249--256 (2010)

\bibitem{gomez2012inferring}
Gomez-Rodriguez, M., Leskovec, J., Krause, A.: Inferring networks of diffusion
  and influence. ACM Transactions on Knowledge Discovery from Data (TKDD)
  \textbf{5}(4),  1--37 (2012)

\bibitem{granovetter1978threshold}
Granovetter, M.: Threshold models of collective behavior. American journal of
  sociology  \textbf{83}(6),  1420--1443 (1978)

\bibitem{guille2012predictive}
Guille, A., Hacid, H.: A predictive model for the temporal dynamics of
  information diffusion in online social networks. In: Proceedings of the 21st
  international conference on World Wide Web. pp. 1145--1152 (2012)

\bibitem{hodas2014simple}
Hodas, N.O., Lerman, K.: The simple rules of social contagion. Scientific
  reports  \textbf{4}, ~4343 (2014)

\bibitem{islam2018deepdiffuse}
Islam, M.R., Muthiah, S., Adhikari, B., Prakash, B.A., Ramakrishnan, N.:
  Deepdiffuse: Predicting the 'who' and 'when' in cascades. In: 2018 IEEE
  International Conference on Data Mining (ICDM). pp. 1055--1060. IEEE (2018)

\bibitem{kempe2003maximizing}
Kempe, D., Kleinberg, J.M., Tardos, {\'{E}}.: Maximizing the spread of
  influence through a social network. In: {KDD'03}. pp. 137--146 (2003)

\bibitem{kingma2014adam}
Kingma, D.P., Ba, J.: Adam: A method for stochastic optimization. arXiv
  preprint arXiv:1412.6980  (2014)

\bibitem{kipf2017semi}
Kipf, T.N., Welling, M.: Semi-supervised classification with graph
  convolutional networks. In: {ICLR'17} (2017)

\bibitem{leskovec2009meme}
Leskovec, J., Backstrom, L., Kleinberg, J.: Meme-tracking and the dynamics of
  the news cycle. In: Proceedings of the 15th ACM SIGKDD international
  conference on Knowledge discovery and data mining. pp. 497--506 (2009)

\bibitem{li2017modeling}
Li, D., Zhang, S., Sun, X., Zhou, H., Li, S., Li, X.: Modeling information
  diffusion over social networks for temporal dynamic prediction. IEEE
  Transactions on Knowledge and Data Engineering  \textbf{29}(9),  1985--1997
  (2017)

\bibitem{liu2018heterogeneous}
Liu, Z., Chen, C., Yang, X., Zhou, J., Li, X., Song, L.: Heterogeneous graph
  neural networks for malicious account detection. In: {CIKM'18}. pp.
  2077--2085. ACM (2018)

\bibitem{mou2016natural}
Mou, L., Men, R., Li, G., Xu, Y., Zhang, L., Yan, R., Jin, Z.: Natural language
  inference by tree-based convolution and heuristic matching. In: {ACL'16}. pp.
  130--136 (2016)

\bibitem{paszke2017automatic}
Paszke, A., Gross, S., Chintala, S., Chanan, G., Yang, E., DeVito, Z., Lin, Z.,
  Desmaison, A., Antiga, L., Lerer, A.: Automatic differentiation in pytorch.
  In: NIPS-W (2017)

\bibitem{Qiu2018deepinf}
Qiu, J., Tang, J., Ma, H., Dong, Y., Wang, K., Tang, J.: Deepinf: Social
  influence prediction with deep learning. In: {KDD'18}. pp. 2110--2119 (2018)

\bibitem{saito2009learning}
Saito, K., Kimura, M., Ohara, K., Motoda, H.: Learning continuous-time
  information diffusion model for social behavioral data analysis. In: Asian
  Conference on Machine Learning. pp. 322--337. Springer (2009)

\bibitem{saito2011learning}
Saito, K., Ohara, K., Yamagishi, Y., Kimura, M., Motoda, H.: Learning diffusion
  probability based on node attributes in social networks. In: International
  Symposium on Methodologies for Intelligent Systems. pp. 153--162. Springer
  (2011)

\bibitem{tambuscio2015fact}
Tambuscio, M., Ruffo, G., Flammini, A., Menczer, F.: Fact-checking effect on
  viral hoaxes: A model of misinformation spread in social networks. In:
  Proceedings of the 24th international conference on World Wide Web. pp.
  977--982 (2015)

\bibitem{vaswani2017attention}
Vaswani, A., Shazeer, N., Parmar, N., Uszkoreit, J., Jones, L., Gomez, A.N.,
  Kaiser, {\L}., Polosukhin, I.: Attention is all you need. In: Advances in
  neural information processing systems. pp. 5998--6008 (2017)

\bibitem{wang2017topological}
Wang, J., Zheng, V.W., Liu, Z., Chang, K.C.C.: Topological recurrent neural
  network for diffusion prediction. In: 2017 IEEE International Conference on
  Data Mining (ICDM). pp. 475--484. IEEE (2017)

\bibitem{wang2017cascade}
Wang, Y., Shen, H., Liu, S., Gao, J., Cheng, X.: Cascade dynamics modeling with
  attention-based recurrent neural network. In: IJCAI. pp. 2985--2991 (2017)

\bibitem{wang2018attention}
Wang, Z., Chen, C., Li, W.: Attention network for information diffusion
  prediction. In: {WWW'18}. pp. 65--66 (2018)

\bibitem{Wang2018SNIDSA}
Wang, Z., Chen, C., Li, W.: A sequential neural information diffusion model
  with structure attention. In: {CIKM'18}. pp. 1795--1798 (2018)

\bibitem{yang2018neural}
Yang, C., Sun, M., Liu, H., Han, S., Liu, Z., Luan, H.: Neural diffusion model
  for microscopic cascade prediction. arXiv preprint arXiv:1812.08933  (2018)

\bibitem{yang2019multi}
Yang, C., Tang, J., Sun, M., Cui, G., Liu, Z.: Multi-scale information
  diffusion prediction with reinforced recurrent networks. In: Proceedings of
  the 28th International Joint Conference on Artificial Intelligence. pp.
  4033--4039. AAAI Press (2019)

\bibitem{yang2010modeling}
Yang, J., Leskovec, J.: Modeling information diffusion in implicit networks.
  In: 2010 IEEE International Conference on Data Mining. pp. 599--608. IEEE
  (2010)

\bibitem{yang2015rain}
Yang, Y., Tang, J., Leung, C.W.k., Sun, Y., Chen, Q., Lit, J., Yang, Q.: Rain:
  social role-aware information diffusion. In: Proceedings of the Twenty-Ninth
  AAAI Conference on Artificial Intelligence. pp. 367--373 (2015)

\bibitem{rumor_yuan_2019}
Yuan, C., Ma, Q., Zhou, W., Han, J., Hu, S.: Jointly embedding the local and
  global relations of heterogeneous graph for rumor detection. In: 2019 IEEE
  International Conference on Data Mining (ICDM). IEEE (2019)

\bibitem{yuan2019learning}
Yuan, C., Zhou, W., Ma, Q., Lv, S., Han, J., Hu, S.: Learning review
  representations from user and product level information for spam detection.
  In: The 19th IEEE International Conference on Data Mining. IEEE (2019)

\bibitem{zhang2018cosine}
Zhang, Y., Lyu, T., Zhang, Y.: Cosine: Community-preserving social network
  embedding from information diffusion cascades. In: Thirty-Second AAAI
  Conference on Artificial Intelligence. pp. 2620--2627 (2018)

\bibitem{zhong2012comsoc}
Zhong, E., Fan, W., Wang, J., Xiao, L., Li, Y.: Comsoc: adaptive transfer of
  user behaviors over composite social network. In: Proceedings of the 18th ACM
  SIGKDD international conference on Knowledge discovery and data mining. pp.
  696--704 (2012)

\end{thebibliography}

\end{document}